\documentclass[runningheads]{llncs}
\usepackage[T1]{fontenc}
\usepackage{graphicx}%
\usepackage{multirow}%
\usepackage{amsmath,amssymb,amsfonts}%
\usepackage[title]{appendix}%
\usepackage{xcolor}%
\usepackage{textcomp}%
\usepackage{manyfoot}%
\usepackage{booktabs}%
\usepackage{listings}%
\usepackage{array}
\usepackage{hyperref}

\begin{document}

\title{Knitting the Ising Transition: Statistical Physics as a Tool for Stochastic Knit Design}

\author{Gabrielle Chang\inst{1} \and 
Kafumi Akaho\inst{2} \and
Tomohiko G. Sano\inst{2,3}\orcidID{0000-0003-0765-8641} \and 
Samuel Poincloux\inst{1}\orcidID{0000-0003-4812-2704}}

\institute{Department of Physical Sciences, Aoyama Gakuin University, 5-10-5 Fuchinobe, Sagamihara, 252-5258, Kanagawa, Japan \and Department of Mechanical Engineering, Faculty of Science and Technology, Keio University, 3-14-1 Hiyoshi, Yokohama, 223-8522, Kanagawa, Japan \and School of Integrated Design Engineering, Graduate School of Science and Technology, Keio University, 3-14-1 Hiyoshi, Yokohama, 223-8522, Kanagawa, Japan}

\maketitle

\begin{abstract}
    Textile designs are typically based on biperiodic, deterministic patterns. However, hand and machine manufacturing are not restricted to periodic patterns. Instead, the entanglement topology can be chosen for each base unit cell of the textile. Non-deterministic processes, such as stochastic realisations of physical systems, can then serve as design tools for creating original patterns. In this Proceedings, we present a proof of concept for stochastic design by generating knitted fabrics from Ising model spin patterns. We leverage the binary analogy between knit/purl stitches and up/down spins to produce fabrics that directly represent spin realisations at different temperatures. Ising patterns then play the role of topological instructions to arrange the entanglements in the fabric. This proof of concept paves the way for novel stochastic designs of fabrics whose properties are directly controlled by the statistical characteristics of an underlying physical process.
    \keywords{knit, Ising model, statistical physics, manufacturing}
\end{abstract}

\section{Introduction}\label{sec1}

Knitted fabrics are planar assemblies formed by a regular entanglement of one or a few yarns into interlocked loops called stitches. 
The stitches cover the plane with a pattern that repeats in two directions, resulting in a doubly periodic entanglement network. 
Thus, a rectangular unit cell with two sets of periodic edges is usually sufficient to describe the structure of a knitted fabric. 
By glueing the corresponding periodic edges together, one finds that the topological nature of a knit is a tangle in a thickened torus~\cite{grishanov2009topologicalI,grishanov2009topologicalII,markande2020knotty,kuzbary2025study}.
In addition to the yarn material, the looped geometry of the stitches and their entanglement pattern control the fabric's properties, particularly mechanical ones, placing knitted fabrics in the category of mechanical metamaterials~\cite{bertoldi2017flexible}. 
For instance, the looped geometry provides stretchiness by harnessing the yarn's soft bending modes rather than rigid stretching~\cite {shanahan1974theoretical,hong2002theoretical,poincloux2018geometry}.
Furthermore, entanglement forces the yarn to bend within the fabric thickness, inducing internal moments that are released through overall fabric curling~\cite {ucar2000prediction,amanatides2022characterizing,tajiri2025curling}. 
Local variations in stitch entanglement and the resulting self-folding tune the stretchiness~\cite{singal2024programming}, control the helical buckling behavior~\cite{Tajiri2026} and sculpt the 3D shape~\cite{niu2025geometric} of the knitted fabrics. 

A single repeated motif is simple to produce and keeps the fabric’s behaviour and appearance predictable and uniform. 
Yet traditional hand-knitting, as well as modern machine-knitting, allows one to control local entanglement patterns within the bounds imposed by knit manufacturing constraints~\cite{shimamoto2026topological}. 
Knitted garment manufacturing has leveraged this design freedom to add stretch or rigidity to specific parts of the fabric, as well as for aesthetic purposes~\cite{emery1995primary}. 
Textiles, and knits in particular, have also recently been identified as promising scaffolds for soft robotics and smart fabrics~\cite{sanchez2021textile}, owing to their highly customisable properties and potential for industrial fabrication. 
Desired local mechanical properties in the fabric are achieved by deterministically varying entanglement, loop geometry, and yarn material~\cite{singal2024programming,dupasquier2025multi}. 
Such knits lose their biperiodicity; their topology is then better represented by meshes~\cite{yuksel2012stitch,wu2019knittable} or directed graph representations~\cite{kapllani2021topoknit,maharaj2025topology}. 

However, both standard biperiodic and locally optimised knits still rely on deterministic patterns. 
Krishma Singal and Sabetta Matsumoto challenged this paradigm by introducing \textit{Statistical Knitting}~\cite{singal2024unraveling}, which leverages randomness and disorder to design fabrics.  
In this article, we propose and demonstrate the feasibility of a stochastic design method based on Ising patterns. 
Proceeding by analogy, we map two stitches of opposite topology, knit and purl, onto up and down spins of the Ising model, a stochastic model of binary spins on a 2D lattice. 
We can thus generate knit and purl patterns from statistical realisations of the Ising model. 
The statistical properties of the patterns are controlled by a single parameter, the effective temperature. 

This proof of concept is organised as follows. 
After detailing the structure of the stitches considered and the fundamentals of the Ising model, we explain how we run the simulations, translate them into knitted patterns, and manufacture the physical samples. 
Based on the resulting fabrics, we comment on how the knitting process and the yarn's frictional and elastic properties transcribe the patterns, and then discuss potential applications of stochastic design.   

\section{Background}\label{sec2}

We begin by introducing the terms and concepts in knitting and statistical physics that we will use throughout this article. 
Via this explanatory background, we also establish the analogy between $\pm1$ Ising spins and knit/purl stitches on which this proof of concept of stochastic knitting is based.

\subsection{Weft knitting and the two fundamental stitches: knit and purl}

\begin{figure}[h]
\centering
\includegraphics[width=1\textwidth]{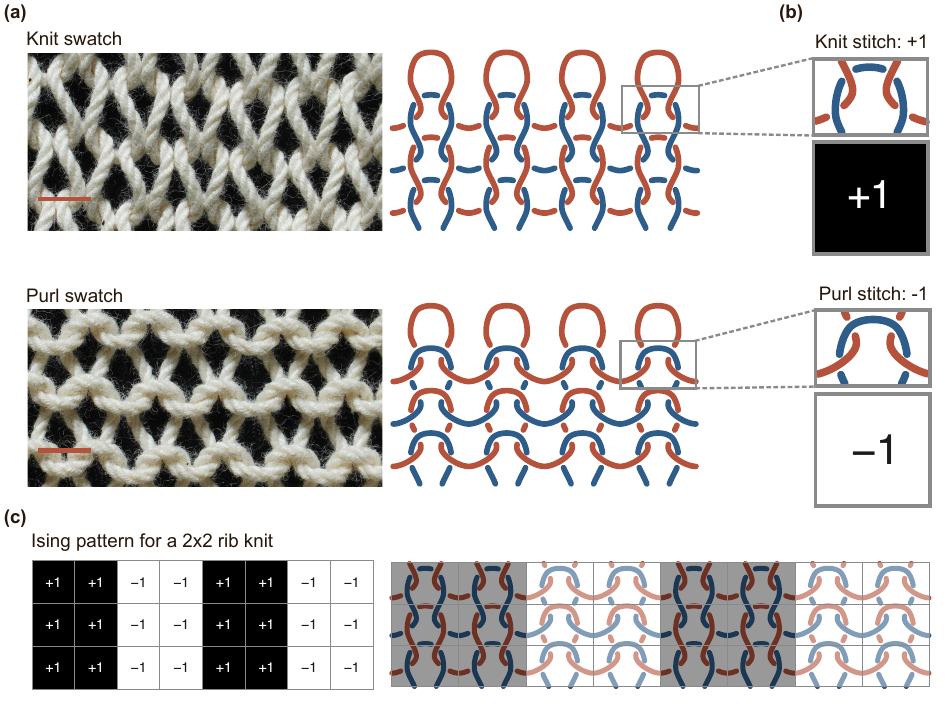}
\caption{(a) Yarn diagrams and pictures of all-knit (top) and all-purl (bottom) swatches. Every crossing is reversed, so the all-purl fabric is the back of the all-knit fabric. Scale bars are both $5\,\mathrm{mm}$. (b) A single knit stitch ($+1$) and purl stitch ($-1$), taken from the boxed regions in (a). (c) 2 × 2 rib, in which two knit columns alternate with two purl columns, shown as a yarn diagram and as a grid of spins (black = $+1$ = knit, white = $-1$ = purl).}\label{fig_1}
\end{figure}

Weft knitting is a type of knitting in which the fabric is built row by row from a continuous yarn. 
Each stitch is a loop pulled through the loop of the row below it, so that the fabric is a lattice of interlocked loops. 
Each stitch is entangled with its neighbours above and below, and the yarn running through it connects it to its left and right neighbours. 
If the same process is repeated for every stitch, the resulting structure is periodic in two directions (Fig.~\ref{fig_1}a).

The fundamental variation consists in pulling the loop from the other side of the fabric. 
The newly created loop thus have the opposite crossing orientation; the yarn that passed over now passes under. 
The two resulting fundamental stitches with opposite crossings (or topology) are called knit and purl. 
When looking at a knitted fabric, knit stitches are recognisable by the loop's legs facing upward and the loop's head downward (Fig.~\ref{fig_1}a, top). 
Inversely, purl stitches have the loop's head facing upward and its legs downward (Fig.~\ref{fig_1}a, bottom). 
Consequently, a swatch of knit stitches becomes a swatch of purl stitches when the fabric is simply viewed from the opposite side.

While many other kinds of stitches exist~\cite{kapllani2021topoknit}, we consider here knitted fabrics composed solely of knit and purl stitches. 
The fabric structure is thus reduced to a rectangular lattice, with each site a binary choice between knit and purl stitches. 
Labelling a knit stitch $+1$ and a purl stitch $-1$ turns any knit/purl pattern into a grid of $\pm 1$ signs, and looking at the same fabric from the other side simply inverts all the signs.  
Figure.~\ref{fig_1}(c) illustrates the corresponding grid for a $2\times 2$ rib pattern. 
A grid of $\pm 1$ is also the way an Ising spin configuration is represented, as explained below. 

\subsection{The Ising model}

The Ising (or Lenz-Ising) model is a canonical model of a phase transition~\cite{lenz1920beitrvsge,ising1925beitrag,brush1967history,goldenfeld2018lectures}, initially introduced as a model for the ferromagnetic transition~\cite{newell1953theory}. 
The model was later developed and applied to many domains, from optimisation problems~\cite{mohseni2022ising} to sociophysics~\cite{mullick2025sociophysics}.
We are interested here in its simplest binary form on a square lattice of spins, where each spin takes one of two values, $+1$ or $-1$. 
Neighbouring spins interact such that aligned spins lower the energy of the system. 
Stochasticity is introduced by placing the system in thermal equilibrium. 
Temperature controls the balance between the energetic preference for alignment and entropy: at low temperatures, ordered configurations dominate, while at high temperatures, spins become increasingly disordered. 
We thus leverage the Ising model as a binary pattern generator that transitions between order and disorder under the control of a single temperature-like parameter. 

We consider a square lattice of size $N$ and a spin configuration $\{s\}$ is given by the set of spin values $s_i=\pm1$, $i\in [1, .., N \times N]$. 
The energy of a configuration is given by the Hamiltonian:
\begin{equation}
H(\{s\}) = -J \sum_{\langle i,j \rangle} s_i s_j \;-\; h \sum_i s_i. \label{eq_Hamiltonian}
\end{equation}
The first term represents interactions between neighbouring spins, controlled by the interaction strength $J>0$.
The sum runs over $\langle i,j \rangle$, all pairs of nearest neighbours on a square lattice. 
As $J > 0$, aligned neighbours ($s_i = s_j$, hence $s_i s_j=+1$) lower the energy and anti-aligned neighbours raise it ($s_i = - s_j$, hence $s_i s_j=-1$). 
Thus, up to an additive constant, the energy is proportional to the total length of the boundaries between up and down domains. 
The second term couples the spins to an external field $h$, which breaks the symmetry by favouring one spin state. 
Throughout this work, we set $h = 0$ to keep the system symmetric. 

At thermal equilibrium at temperature T, the probability of each configuration is given by the Boltzmann distribution:
\begin{equation}
    P(\{s\}) = \frac{1}{Z}\, e^{-H(\{s\})/k_B T}, \qquad Z = \sum_{\{s\}} e^{-H(\{s\})/k_B T}, \label{eq_Boltzmann}
\end{equation}
with $k_B$ the Boltzmann constant and $Z$ the partition function of the configurations~\cite{goldenfeld2018lectures}. 
Without external fields, the only relevant physical quantity is $J/k_B$, which we set equal to $1$. 

Two observables are computed from a configuration $\{s\}$, the average energy:
\begin{equation}
    e = \frac{H(\{s\})}{N^2}, \label{eq_energy}
\end{equation}
and the average magnetisation:
\begin{equation}
    m = \frac{1}{N^2} \sum_i s_i,. \label{eq_mag}
\end{equation}

Without external fields ($h=0$), the average energy is a measure of the density of domain walls and smoothly increases with temperature.  
The average magnetisation measures the net imbalance between up and down spins. 
It slowly decreases with temperature, before falling steeply to $0$ at a critical temperature $T=T_c$. 
The magnetisation is the order parameter for the order/disorder phase transition. 

On the infinite two-dimensional ($N\rightarrow\infty$) square lattice, Lars Onsager provided an exact solution for the critical temperature defined by $\sinh(2J/k_B T_c) = 1$~\cite{onsager1944crystal,baxter2011onsager}, that is,
\begin{equation}
    T_c = \frac{2J}{k_B \ln\!\left(1+\sqrt{2}\right)} \approx 2.269. \label{eq_crit_temp}
\end{equation}
Near $T_c$, regions of aligned spins (domains) appear on all length scales. 
In a finite-size system featuring boundary conditions (such as periodic or Dirichlet-type value-imposed conditions), the transition is less sharp, but its nature is not altered, nor is the value of $T_c$.

\subsection{From spins to stitches}

We use the Ising model purely as a pattern-generation system. 
It provides a family of binary patterns on a square lattice, controlled by a single parameter $k_BT/J$, and with statistical properties given by analytical formulas. 
We sample configurations at several temperatures to see how the transition from order to disorder appears in textile patterns.

The mapping to knitting is direct. 
Each spin configuration is represented as a two-dimensional grid whose two states can be mapped to knitting instructions: spin $+1$ becomes a knit stitch and spin $-1$ a purl stitch. 
The Ising patterns thus provide topological instructions for organising the crossings in the fabric. 
The model's global spin-flip symmetry has a physical counterpart because a purl stitch is a knit stitch seen from behind.
Flipping every spin with a left-right reflection is the same as turning the fabric over. 
Since both are symmetries of the model at $h=0$, the back of each swatch (or configuration $\{s\}$) is an equally probable configuration with the same energy but the opposite magnetisation $-m$.
In the following, we detail all steps from simulating Ising model realisations to sample fabrication. 

\section{Methods}

Our workflow has four stages: simulate spin configurations, convert them to two-colour stitch images with a border and padding, knit the images on a digital knitting machine, and post-process the swatches.

\subsection{Ising simulations}

We implemented the two-dimensional Ising model simulations in Python on an $N\times N$ square lattice with periodic boundary conditions, so that every site has exactly four neighbours and the pattern has no artificial edges. 
The overall numerical scheme to reach a pattern for a given temperature is to start with a randomly generated spin configuration and then iteratively select one or several spins, flip their sign, and accept the change if it lowers the energy. If the flip increases the energy, it can still be accepted with a probability given by the Boltzmann weight associated with the energy change induced by the flip. 
Thus, high-temperature configurations are more likely to accept energy-increasing flips, leading to disordered configurations. 
The accept/reject flipping process is performed until the energy of the configuration converges. 
Since flipping all the spins does not change the energy, a configuration with $|m|>0$ has an equal probability of having dominant $+1$ spins ($m>0$) or $-1$ spins ($m<0$). 

\begin{figure}[h]
\centering
\includegraphics[width=1\textwidth]{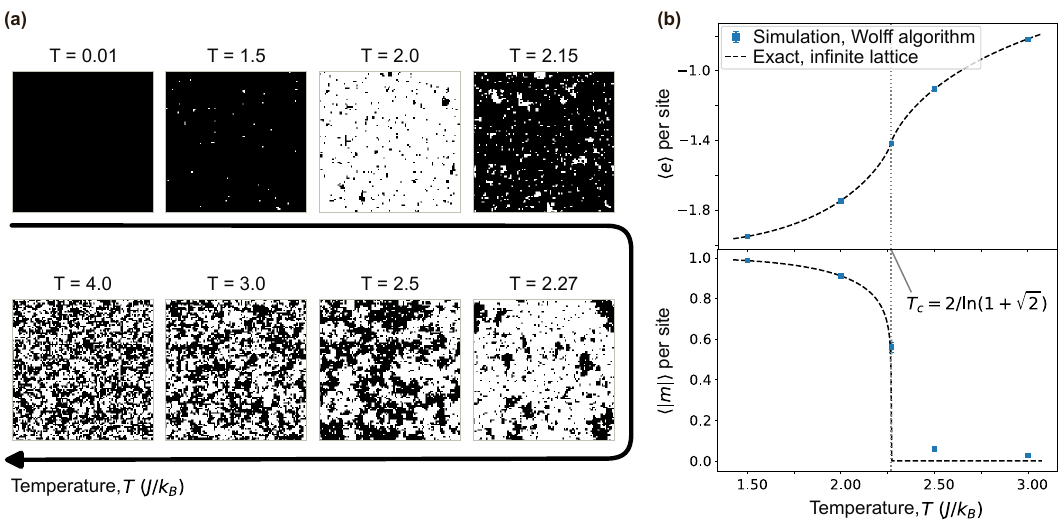}
\caption{(a) Ising configurations from Wolff simulations at eight temperatures, read along the arrow from ordered ($T = 0.01$) to disordered ($T = 4.0$); black = $+1$ = knit, white = $-1$ = purl. For these realisations, only $T = 2.0$ has $m<0$ (dominant white $-1$ spins). The configuration with $T=2.27>T_c$ retains non-zero magnetisation due to the finite lattice size. 
(b) Energy per site $\langle e\rangle$ and magnetisation ⟨$|m|$⟩ against temperature from simulation (points), with the exact infinite-lattice results (dashed)~\cite{onsager1944crystal,yang1952spontaneous}. The dotted line marks $T_c=2J/(k_B \ln(1+\sqrt{2}))\approx 2.269$. Above $T_c$, finite-size effects keep $\langle|m|\rangle$ slightly above zero.}\label{fig_2}
\end{figure}

The standard method uses the Metropolis algorithm~\cite{metropolis1953equation}, which flips one spin at a time. 
However, this algorithm easily gets trapped in metastable states at low energy (such as stripes spanning the lattice) and suffers acute slowing down near the critical temperature. 
We thus sampled configurations using the Wolff cluster algorithm~\cite{wolff1989collective}. 
The Wolff algorithm instead grows a cluster of aligned spins and flips it in a single move, so it can escape metastable states and remains efficient near $T_c$. 
Figure~\ref {fig_2} shows a series of resulting Ising configurations of size $N=100$ at increasing temperatures. 
To verify the validity of the simulations, we compared the computed average energy $\langle e\rangle$ and magnetisation $|m|$ with their analytical predictions for an infinite lattice~\cite{onsager1944crystal,yang1952spontaneous}. 
The simulations agreed quantitatively with the predictions (Fig.~\ref{fig_2}), except for the magnetisation near the critical temperature $T_c$ due to the finite size of the simulated lattice.

Simulations at each temperature and realisation were run independently and parallelised across CPU cores. 
We used NumPy and Numba for simulation, and Pandas, Matplotlib, and Pillow for analysis, visualisation, and image processing. 
The knitted patterns were generated with version $0.1.0$ of the repository (tag $v0.1.0$), and the exact image files that were knitted are archived in its sample\_knits folder. 
The simulation and image-processing code is available in the following repository~\cite{github}. 

\subsection{From Ising to machine instructions}

\begin{figure}[h]
\centering
\includegraphics[width=1\textwidth]{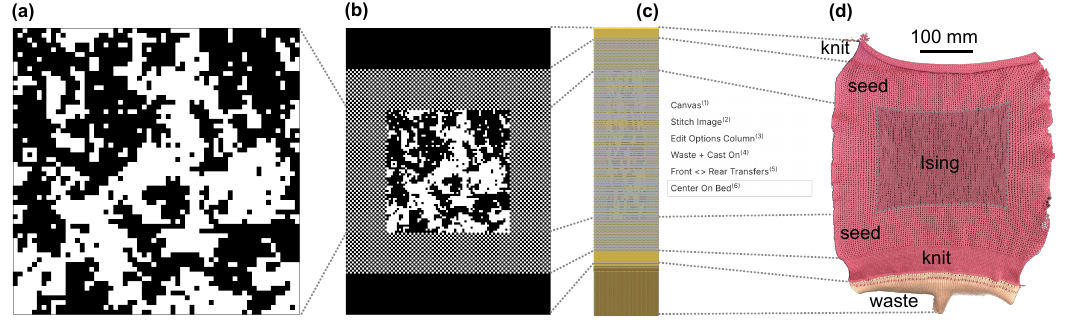}
\caption{From simulated configuration to knit swatch, shown for an early test sample. (a) A $N=60$ Ising configuration from a Wolff simulation; black = $+1$ = knit, white = $-1$ = purl. (b) The configuration padded with a 20-stitch checkerboard border and 20 rows of plain knit above and below. (c) The image in the Kniterate editor with the set of editor commands. (d) The knitted swatch (stitch size 8), with the waste yarn in light pink at the bottom. Dotted grey lines follow the vertical positions of the different parts across the steps.}\label{fig_3}
\end{figure}

Spin configurations were saved as a binary image of $N\times N$ pixels (Fig.~\ref{fig_3}a). 
Each pixel represents a stitch: black for spin $+1$ (knit) and white for spin $-1$ (purl).
The knitting process adds physical boundaries to the pattern that cannot reproduce the periodic boundary conditions of the simulations.
We thus add artificial boundaries to surround the spin pattern with fabrics of known mechanical properties.
Because internal moments from the combination of over-under crossings and yarn elasticity create intrinsic curvatures, stitches force the fabric to curl and take on 3D shapes~\cite{niu2025geometric,tajiri2025curling}. 
By symmetry, knit and purl stitches have opposite intrinsic curvatures.
To observe the Ising patterns, we chose boundaries that suppress curling at the fabric edges. 
By alternating in a checkerboard pattern knit ($+1$) and purl ($-1$) stitches, the overall intrinsic curvature averages to zero and the fabric remains flat. 
We then padded the Ising pattern with a $20$-stitch-wide border of this checkerboard pattern (also called seed stitch). 
Completely surrounding the Ising pattern with this flat border constrains the resulting knit to stay mostly 2D, preventing it from self-folding into complex 3D shapes.
Finally, we added an extra $20$ rows of plain knit above and below the bordered pattern. 
These rows separate the patterned region from the waste yarn at the start and end of knitting, making the transition between stitch types easier for the machine. 
This padding is generated with the simulation code by directly increasing the image size to $(N+n_w)\times(N+n_c)$, where the extra wale and course knitting numbers are set as $(n_w, n_c) = (40, 80)$ throughout, and keeping the same convention with the $+1/-1$ corresponding to the knit and purl stitches (Fig.~\ref{fig_3}b).

Next, we translated the binary image into sets of knitting-machine instructions. 
We used a Kniterate knitting machine with two beds of hook-like needles. 
The machine knits row by row, with a carriage travelling over the beds, moving the needles and feeding yarn according to programmed instructions.
The images were converted to machine instructions in the browser-based Kniterate editor by building a stack of editor commands on top of the image (Fig.~\ref{fig_3}c):
\begin{itemize}
\item The command \textit{Stitch Image} imports two-colour images and maps each colour to a stitch type. 
Black was mapped to FRNT\_KNIT (a knit on the front bed) and white to REAR\_KNIT (a knit on the rear bed, which appears as a purl from the front). 
\item \textit{Edit: Options Column} sets key practical parameters, such as stitch size, carriage speed, and tension, from the pull-down rollers. 
\item Then, \textit{Waste plus Cast On} adds waste yarn at the bottom of the pattern. 
Knits are fabricated starting from the first row at the bottom, and waste yarn is a preset sequence of stitches that helps start the work and apply the right tension at the bottom of the knit to execute the instructions correctly. 
As its name suggests, it is meant to be removed after completion. 
\item The \textit{Front $<>$ Rear Transfers} command detects where a stitch changes bed from one row to the next and inserts the transfers. 
A transfer moves a loop from one needle to another, for instance between the front and rear beds. 
A stitch to be purled is moved to the rear bed and knitted there in the next row, then moved back to the front bed when it becomes a knit again. 
\item Finally, \textit{Center On Bed} centres the design on the needle bed.
The editor’s check was run before export. 
The design was exported as a .kc file, renamed command.kc (the only file name the machine reads), and copied to an SD card. 
The card was inserted into the machine before it was switched on, and knitting was started from the machine’s menu. 
After completion, the knitted fabric drops from the bottom of the machine (Fig.~\ref{fig_3}d).
\end{itemize}

\subsection{Physical fabrication and post-processing}

We knitted swatches on a Kniterate digital flatbed knitting machine (7 gauge) using 3-ply cotton yarn from SAORI. 
Table~\ref{tab_1} shows the detailed machine settings.
We selected the settings based on several test swatches. 
In some samples, larger stitch sizes left large gaps in the patterned region, which we attribute to extra yarn drawn by front-to-rear transfers. 
Reducing the stitch size or increasing tension on every row produced holes or broke the yarn, so we reduced the stitch size to 7 on knitting rows only, leaving the transfer rows unchanged. 
The carriage speed was also reduced from the editor default, because slower speeds produced fewer knitting errors. 
Holes also tended to appear where the fabric passed from the waste yarn to the main knit.
Overall, knitting required close supervision. 
The machine often stopped on clump detection even when no clump was visible, and operators had to clear each stop by hand. Additionally, broken needles caused dropped stitches and holes. 
Whenever a swatch had a defect, we restarted it from the beginning.
Because of the many transfers, low carriage speed, and clump errors, each swatch took several hours of machine time to complete. 

\begin{table}[ht]
\centering
\begin{tabular}{|p{0.35\textwidth} | p{0.55\textwidth}|}
\hline
\textbf{Setting} & \textbf{Value} \\ \hline
Machine    & Kniterate digital flatbed, 2 beds $\times$ 252 needles, 7 needles per inch; firmware: v.1.0.14    \\ \hline

Yarn    & SAORI 3-ply cotton, diameter $\approx 1\,\mathrm{mm}$,  linear density $\approx 0.18\,\mathrm{g/m}$   \\ \hline

Editor    & Kniterate editor (editor.kniterate.design), browser-based (August 2025)  \\ \hline

Knitting image    & 140 stitches $\times$ 180 rows per sample   \\ \hline

color mapping (Stitch Image v1.0.0)    & Black $(\#000000)$ $\rightarrow$ FRNT\_KNIT (knit); white $(\#ffffff)$ $\rightarrow$ REAR\_KNIT (purl); single pass   \\ \hline

Knitting rows (Edit Options Column v1.0.1)   & Stitch size 7; carriage speed: 200 mm/sec; roller pull-down: editor defaults   \\ \hline

Transfer rows (Front $<>$ Rear Transfers v1.0.0, between rows)    & Speed 120, roller 0, stitch size 0   \\ \hline

Yarn carriers    & Body yarn: 3; waste yarn: 1; draw thread: 6   \\ \hline

Waste + Cast On (v1.0.0)    & 80 interlock rows; all other options at defaults   \\ \hline

Placement    & Center On Bed (global)   \\ \hline

Compiler settings    & Carrier spacing 2; stopping distance 2.5; maximum float distance 7   \\ \hline

Machine file    & Exported .kc file renamed command.kc, loaded from an SD card   \\
\hline
\end{tabular}
\vspace{2mm}
\caption{Machine and editor settings used for the final version ($N=100$) of the samples. Settings not listed were left at the editor’s defaults. \label{tab_1}}
\end{table}

\begin{figure}[h]
\centering
\includegraphics[width=1\textwidth]{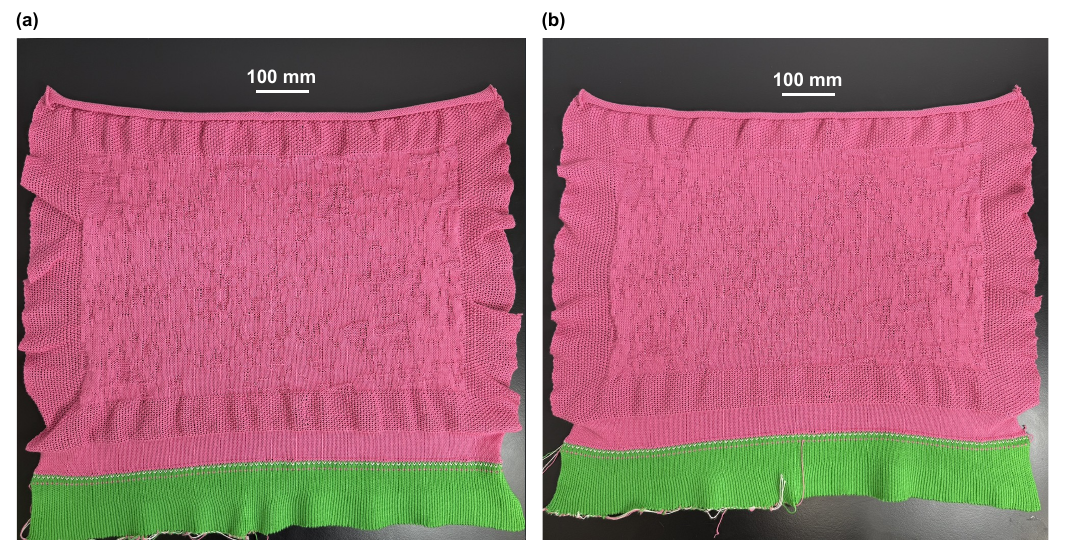}
\caption{A knitted swatch (a) as taken off the machine and (b) after blocking. Blocking relaxes the fabric: the stitches in the patterned region are more even, and the diagonal distortions from the take-down tension are reduced. The green band at the bottom is waste yarn, removed after blocking.}\label{fig_4}
\end{figure}

Immediately after removal from the Kniterate machine, swatches are distorted by take-down tension, uneven curling of knit and purl regions, and apparent holes due to minute variations in yarn length per stitch. 
In both handknitting and machine manufacturing, post-processing steps are standard to obtain fabrics with reproducible properties, appearance and shape. 
Blocking is a common post-processing step in which the fabric is wetted or steamed and dried while held at a fixed shape so that the yarn relaxes. 
Each swatch was first stretched by hand in both directions for approximately one minute, then wetted and dried with a hair dryer while gently pulled-taut for about 10 minutes before air drying fully. 
After blocking, we cut off the waste yarn. 
Figure~\ref{fig_4} shows a typical sample before and after blocking. 
For our samples, blocking did not produce a dramatic change, but we still observed more even patterns and less distortion between the different knitted regions.  
Other yarns and finishing procedures may produce different or more pronounced effects. 
The final preparation step involves laying the samples flat and holding them under slight tension with laser-cut acrylic plates. 
This presentation prevents any free 3D self-folding of the fabric to maximise visibility of the resulting patterns.

\section{Results and Discussion}

In this section, the proof of concept of stochastic design is demonstrated by manufacturing knits according to the previously detailed methods. 
We then comment on how the patterns are expressed in the resulting fabrics, as well as limitations and future prospects of Ising-based stochastic designs.

\subsection{Knitting across the Ising transition}

\begin{figure}[h!]
\centering
\includegraphics[width=1\textwidth]{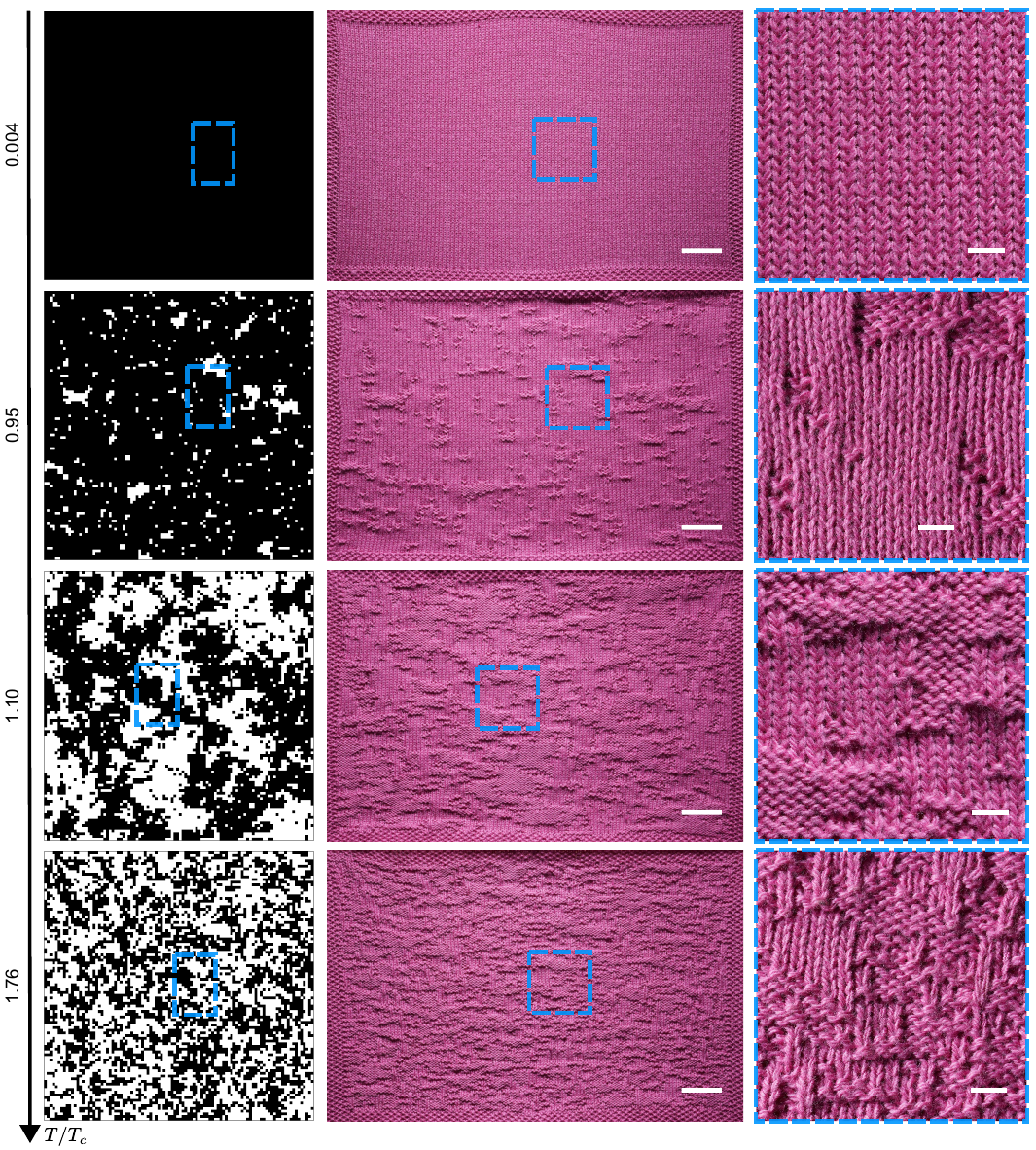}
\caption{Left: Generated Ising configurations at $T/T_c=0.004, 0.95, 1.10, 1.76$. Middle: Corresponding front of each physical swatch, held flat and slightly stretched by an acrylic frame. White scale bar: $30\,\mathrm{mm}$. Right: Close-up view of the knitted patterns. For each temperature, the zoomed region is highlighted by a blue rectangle on the corresponding knit and pattern. White scale bar: $5\,\mathrm{mm}$.}\label{fig_5}
\end{figure}

We selected $N = 100$ Ising patterns at four different temperatures across the phase transition $T=0.01, 2.15, 2.50, 4.00$. 
These temperature span the three regimes of the model, with $T/T_c \approx$ $0.004$ (ordered), $0.95$ and $1.10$ (just below and above $T_c$), and $ 1.76$ (disordered).
At each temperature, we generated ten configurations and hand-picked the one we physically fabricated. 
For consistency, for patterns where one stitch dominates, we picked knit(+1)-dominated ones. 
Figure~\ref {fig_5} shows the four knitted and blocked patterns, and the corresponding simulated configurations and high-resolution photographs of the samples are available in a Zenodo repository~\cite{zenodo}.
The combined effects of yarn elasticity, friction and post-processing express the crossing topology given by the Ising pattern into a shaped object. 

We now comment on the result of this physical interpretation of the patterns. 
First, the patterns are free of machine and instruction errors; the binary patterns are then reproduced strictly as knit and purl stitches.
The Ising patterns remain well recognisable on the knits, with knit and purl regions appearing in low-relief within the fabric. 
This low-relief effect modifies the fabric's appearance, texture, and tactile feel, and evolves with temperature.
\begin{itemize}
    \item At low temperature ($T/T_c = 0.004$), the swatch is a uniform sea of knits. 
    \item At $T/T_c = 0.95$, small, scattered purl islands emerge and stand out from the sea of knits. 
    \item At $T/T_c = 1.10$, akin to land and water in a half-flooded marsh, knit and purl stitches cover similar areas in irregular and interlocking domains of many sizes.  
    \item At the highest temperature $T/T_c = 1.76$, the domains shrink to a fine, pebbled texture. 
\end{itemize}

Additionally, the mechanical interpretation of the patterns by the yarns introduces anisotropy. 
The patterns lose their square shape and become rectangular, with an aspect ratio where the width exceeds the height.
This anisotropy directly reflects stress relaxation in the knits, which favours less-bent yarns by increasing stitch width~\cite {munden1959geometry,choi2003energy,crassous2024metastability}. 
The resulting aspect ratio appears insensitive to the temperature. 

\subsection{Current limitations and future prospects}

This work aimed to demonstrate a proof of concept for stochastic design of knitted fabrics. 
We leveraged the analogy between the up/down spins of the Ising model and the opposite topology of knit/purl stitches as a design tool.
An interesting direction would be to assess how far we can push the analogy. 
For instance, how frustration of opposite-sign spins increases a configuration's energy could relate to how arrangements of stitches with opposite topology frustrate the fabric by preventing intrinsic curvature relaxation. 
Similarly, signs of the phase transition, noticeable in the magnetisation of the Ising model, could have a counterpart in an averaged physical property of the knitted fabrics. 
A small change in a parameter could then induce drastic changes in textile properties. 
Even without the sharpness of a phase transition, stochastic design could be a powerful tool for continuously and finely controlling the physical properties of fabrics, such as their geometry and mechanics.

Periodic designs offer a set of discrete responses. 
For example, stockinette and rib differ strongly in how easily they stretch across their width~\cite{singal2024programming}. 
A target response between the two is hard to reach with a few motifs, whereas a stochastic pattern family controlled by continuous parameters could move between them gradually. 
The Ising model has two such parameters: the field $h$ sets the knit fraction, and the temperature $T$ controls the density and size of knit–purl domains. 
If a property such as stiffness depends mainly on the knit fraction, it would vary smoothly with $h$, as the magnetisation does. 
If it depends mainly on the boundaries between domains, it would track the energy and vary with $T$.

However, the choice of the statistical model may impose practical limitations. 
In the case of the Ising model, the inherent noisy patterns require many yarn transfers between needles, increasing fabrication time and error probabilities. 
This limitation depends strongly on temperature for the Ising model (the $T = 4.0$ swatch needs about $35\%$ more transfers than the $T = 0.01$ swatch), but also on the choice of the padding borders (the checkerboard border alone requires thousands of transfers, because every border stitch changes type in every row). 
By exploring different boundaries or statistical models, one could target more manufacturing-friendly designs or different mechanical properties. 
Stochastic waveforms combined with procedural graphic techniques have already been used to create original woven designs that mix traditional weaving with digital fabrication~\cite{hayama2025wave2weave}.  
The platform explored by Krishma Singal and Elisabetta Matsumoto uses knits whose knit/purl patterns are generated by the Cahn-Hilliard equation~\cite{cahn1958free}.
This equation describes the phase transition by spinodal decomposition between two components. 
It produces smooth, coarsening domains at a fixed knit fraction instead of the noisy boundaries of Ising patterns. 
These smooth bounded domains seed self-folding that, in turn, affects the overall shape and stretching properties of the fabrics, achieving smooth stochastic control~\cite{singal2024unraveling,matsumoto2026statistical}. 

\section{Conclusion}

We challenge the notion of textiles as periodic tangles by introducing stochastic Ising knit design. 
Relying on the binary analogy between the opposite signs of up/down spins in an Ising model and the opposite topology of knit/purl stitches in a knitted fabric, we turned equilibrium spin configurations into machine-knittable patterns. 
By varying the equilibrium temperature across an order/disorder phase transition, we explored the aesthetic and textural properties of the resulting physical realisations. 
The low-relief texture of manufactured knits ranges from smooth, uniform stockinette to fine, random roughness, passing through scale-free, interlocked reliefs near the critical temperature. 
Such stochastic designs could support outreach by combining visual and tactile sensations to convey concepts from statistical physics and critical phenomena. 
Beyond this proof of concept, we envision stochastic patterns that fully exploit the design space of digital and manual knitting as a route to continuously control textile properties with only a handful of parameters. 

\textbf{Acknowledgements}
G.C. was supported by the Thomas J. Watson Fellowship and thanks members of the SIGMA Lab and the Friction Physics Lab for discussions, as well as the Kniterate Engineering support team for their support. T.G.S. was supported by MEXT KAKENHI 24H00299 and JST FOREST Program, Grant Number JPMJFR212W. S.P. thanks H. Matsukawa for fruitful discussions and administrative support, and acknowledges the financial support provided by JSPS KAKENHI Grant-in-Aid for Early-Career Scientists No. 25K17363. 

\textbf{Data and code availability}
Simulation and image-processing code is made available (\href{https://github.com/gabbiechang/knit-ising-model}{github.com/gabbiechang/knit-ising-model}). The swatches reported here were generated with release v0.1.0, archived on Zenodo (\href{https://doi.org/10.5281/zenodo.23552182}{doi:10.5281/zenodo.23552182}) together with high-resolution photographs of the knitted samples. Further simulation outputs and Kniterate notes are available from the authors on request.

\textbf{Disclosure of Interests}
The authors have no competing interests to declare that are relevant to the content of this article.

\bibliographystyle{splncs04}
\bibliography{knit.bib}

@article{poincloux2018geometry,
  title={Geometry and elasticity of a knitted fabric},
  author={Poincloux, Samuel and Adda-Bedia, Mokhtar and Lechenault, Fr{\'e}d{\'e}ric},
  journal={{\it Phys. Rev. X}},
  volume={8},
  number={2},
  pages={021075},
  year={2018},
  publisher={APS}
}

@article{crassous2024metastability,
  title={Metastability of a Periodic Network of Threads: Shapes of a Knitted Fabric},
  author={Crassous, J{\'e}r{\^o}me and Poincloux, Samuel and Steinberger, Audrey},
  journal={{\it Phys. Rev. Lett.}},
  volume={133},
  number={24},
  pages={248201},
  year={2024},
  publisher={APS}
}

@article{tajiri2025curling,
  title={Curling morphology of knitted fabrics: Structure and Mechanics},
  author={Tajiri, Kotone and Murakami, Riki and Kobayashi, Shunsuke and Tarumi, Ryuichi and Sano, Tomohiko G},
  journal={{\it Extreme Mech. Lett.}},
  volume={76},
  pages={102300},
  year={2025},
  publisher={Elsevier}
}

@article{Tajiri2026,
  title     = {Buckling of knitted fabric wrapped around a rigid cylinder},
  author    = {Tajiri, Kotone and Sano, Tomohiko G},
  journal   = {J. Appl. Mech.},
  volume    = {93},
  pages     = {051004},
  year      = {2026},
  publisher = {American Society of Mechanical Engineers},
  doi       = {10.1115/1.4071490}
}

@article{singal2024programming,
  title={Programming mechanics in knitted materials, stitch by stitch},
  author={Singal, Krishma and Dimitriyev, Michael S and Gonzalez, Sarah E and Cachine, A Patrick and Quinn, Sam and Matsumoto, Elisabetta A},
  journal={{\it Nat. Commun.}},
  volume={15},
  number={1},
  pages={2622},
  year={2024},
  publisher={Nature Publishing Group UK London}
}

@article{grishanov2009topologicalI,
  title={A topological study of textile structures. Part I: An introduction to topological methods},
  author={Grishanov, Sergei and Meshkov, Vadim and Omelchenko, Alexander},
  journal={{\it Text. Res. J.}},
  volume={79},
  number={8},
  pages={702--713},
  year={2009},
  publisher={Sage Publications Sage UK: London, England}
}

@article{grishanov2009topologicalII,
  title={A topological study of textile structures. Part II: Topological invariants in application to textile structures},
  author={Grishanov, S and Meshkov, Vadim and Omelchenko, A},
  journal={{\it Text. Res. J.}},
  volume={79},
  number={9},
  pages={822--836},
  year={2009},
  publisher={Sage Publications Sage UK: London, England}
}

@article{niu2025geometric,
  title={Geometric modeling of knitted fabrics},
  author={Niu, Lauren and Dion, Genevi{\`e}ve and Kamien, Randall D},
  journal={{\it Proc. Natl. Acad. Sci. USA}},
  volume={122},
  number={7},
  pages={e2416536122},
  year={2025},
  publisher={National Academy of Sciences}
}

@book{goldenfeld2018lectures,
  title={Lectures on phase transitions and the renormalization group},
  author={Goldenfeld, Nigel},
  year={2018},
  publisher={CRC Press}
}

@book{emery1995primary,
  title={The primary structures of fabrics: an illustrated classification.},
  author={Emery, Irene},
  year={1995},
  publisher={Whitney Library of Design}
}

@article{yuksel2012stitch,
  title={Stitch meshes for modeling knitted clothing with yarn-level detail},
  author={Yuksel, Cem and Kaldor, Jonathan M and James, Doug L and Marschner, Steve},
  journal={ACM Transactions on Graphics (TOG)},
  volume={31},
  number={4},
  pages={1--12},
  year={2012},
  publisher={ACM New York, NY, USA}
}

@article{wu2019knittable,
  title={Knittable stitch meshes},
  author={Wu, Kui and Swan, Hannah and Yuksel, Cem},
  journal={ACM Transactions on Graphics (TOG)},
  volume={38},
  number={1},
  pages={1--13},
  year={2019},
  publisher={ACM New York, NY, USA}
}

@article{kapllani2021topoknit,
  title={TopoKnit: a process-oriented representation for modeling the topology of yarns in weft-knitted textiles},
  author={Kapllani, Levi and Amanatides, Chelsea and Dion, Genevieve and Shapiro, Vadim and Breen, David E},
  journal={Graphical models},
  volume={118},
  pages={101114},
  year={2021},
  publisher={Elsevier}
}

@article{maharaj2025topology,
  title={Topology, integrity, and stability analysis of weft-knitted textiles},
  author={Maharaj, Levi Kapllani and Amanatides, Chelsea and Dion, Genevieve and Shapiro, Vadim and Breen, David E},
  journal={Textile Research Journal},
  pages={00405175241306133},
  year={2025},
  publisher={SAGE Publications Sage UK: London, England}
}

@misc{kuzbary2025study,
      title={A study of 2-periodic weft-knitted textiles using the theory of knots and links}, 
      author={Miriam Kuzbary and Shashank G. Markande and Elisabetta A. Matsumoto and Stanley Pritchard},
      year={2025},
      eprint={2510.08384},
      archivePrefix={arXiv},
      primaryClass={math.GT},
      url={https://arxiv.org/abs/2510.08384}, 
}

@inproceedings{markande2020knotty,
  author      = {Markande, Shashank G and Matsumoto, Elisabetta},
  title       = {Knotty Knits are Tangles in Tori},
  pages       = {103--112},
  booktitle   = {Proceedings of Bridges 2020: Mathematics, Art, Music, Architecture, Education, Culture},
  year        = {2020},
  editor      = {Yackel, Carolyn and Bosch, Robert and Torrence, Eve and Fenyvesi, Krist\'{o}f},
  isbn        = {978-1-938664-36-6},
  issn        = {1099-6702},
  publisher   = {Tessellations Publishing},
  address     = {Phoenix, Arizona},

  url         = {http://archive.bridgesmathart.org/2020/bridges2020-103.html}
}

@article{bertoldi2017flexible,
  title={Flexible mechanical metamaterials},
  author={Bertoldi, Katia and Vitelli, Vincenzo and Christensen, Johan and Van Hecke, Martin},
  journal={Nature Reviews Materials},
  volume={2},
  number={11},
  pages={1--11},
  year={2017},
  publisher={Nature Publishing Group}
}

@article{hong2002theoretical,
  title={Theoretical analysis of load-extension properties of plain weft knits made from high performance yarns for composite reinforcement},
  author={Hong, H and De Araujo, MD and Fangueiro, R and Ciobanu, O},
  journal={Textile research journal},
  volume={72},
  number={11},
  pages={991--996},
  year={2002},
  publisher={Sage Publications, Sage CA: Thousand Oaks, CA}
}

@article{shanahan1974theoretical,
  title={A theoretical analysis of the tensile properties of plain-knitted fabrics. part i: The load-extension curve for fabric extension parallel to the courses},
  author={Shanahan, WJ and Postle, R},
  journal={Journal of the Textile Institute},
  volume={65},
  number={4},
  pages={200--212},
  year={1974},
  publisher={Taylor \& Francis}
}

@article{amanatides2022characterizing,
  title={Characterizing and predicting the self-folding behavior of weft-knit fabrics},
  author={Amanatides, Chelsea and Ghita, Oana and Evans, Ken E and Dion, Genevieve},
  journal={Textile Research Journal},
  volume={92},
  number={21-22},
  pages={4060--4076},
  year={2022},
  publisher={SAGE Publications, Sage UK: London, England}
}

@article{ucar2000prediction,
  title={Prediction of Curling Distance of Dry-Relaxed Cotton Plain Knitted Fabrics Part 1: Theoretical analysis of moments on the loop that force the fabric to curl},
  author={Ucar, Nuray},
  journal={Journal of Textile Engineering},
  volume={46},
  number={4},
  pages={109--117},
  year={2000},
  publisher={The Textile Machinery Society of Japan}
}

@article{shimamoto2026topological,
  title={Topological Defect Propagation to Classify Knitted Fabrics},
  author={Shimamoto, Daisuke S and Shimamoto, Keiko and Mahmoudi, Sonia and Poincloux, Samuel},
  journal={Physical Review X},
  volume={16},
  number={3},
  pages={031006},
  year={2026},
  publisher={APS}
}

@article{sanchez2021textile,
author = {Sanchez, Vanessa and Walsh, Conor J. and Wood, Robert J.},
title = {Textile Technology for Soft Robotic and Autonomous Garments},
journal = {Advanced Functional Materials},
volume = {31},
number = {6},
pages = {2008278},
doi = {https://doi.org/10.1002/adfm.202008278},
year = {2021}
}

@article{dupasquier2025multi,
  title={Multi-level mechanical modeling and computational design framework for weft knitted fabrics},
  author={du Pasquier, Cosima and Jeong, Sehui and Liu, Pan and Williams, Susan and Mnejja, Nour and Okamura, Allison M and Tibbits, Skylar and Chen, Tian},
  journal={Extreme Mechanics Letters},
  pages={102423},
  year={2025},
  publisher={Elsevier}
}

@article{lenz1920beitrvsge,
  title={Beitr{\v{s}}ge zum verst{\v{s}}ndnis der magnetischen eigenschaften in festen k{\v{s}}rpern},
  author={Lenz, Wilhelm},
  journal={Physikalische Z},
  volume={21},
  number={613-615},
  pages={1},
  year={1920}
}

@article{ising1925beitrag,
  title={Beitrag zur theorie des ferromagnetismus},
  author={Ising, Ernst},
  journal={Zeitschrift f{\"u}r Physik},
  volume={31},
  number={1},
  pages={253--258},
  year={1925},
  publisher={Springer}
}

@article{onsager1944crystal,
  title={Crystal statistics. I. A two-dimensional model with an order-disorder transition},
  author={Onsager, Lars},
  journal={Physical review},
  volume={65},
  number={3-4},
  pages={117},
  year={1944},
  publisher={APS}
}

@article{brush1967history,
  title={History of the Lenz-Ising model},
  author={Brush, Stephen G},
  journal={Reviews of modern physics},
  volume={39},
  number={4},
  pages={883},
  year={1967},
  publisher={APS}
}

@article{newell1953theory,
  title={On the theory of the Ising model of ferromagnetism},
  author={Newell, Gordon F and Montroll, Elliott W},
  journal={Reviews of Modern Physics},
  volume={25},
  number={2},
  pages={353},
  year={1953},
  publisher={APS}
}

@article{mohseni2022ising,
  title={Ising machines as hardware solvers of combinatorial optimization problems},
  author={Mohseni, Naeimeh and McMahon, Peter L and Byrnes, Tim},
  journal={Nature Reviews Physics},
  volume={4},
  number={6},
  pages={363--379},
  year={2022},
  publisher={Nature Publishing Group UK, London}
}

@article{mullick2025sociophysics,
  title={Sociophysics models inspired by the Ising model},
  author={Mullick, Pratik and Sen, Parongama},
  journal={The European Physical Journal B},
  volume={98},
  number={9},
  pages={206},
  year={2025},
  publisher={Springer}
}

@article{baxter2011onsager,
  title={Onsager and Kaufman’s calculation of the spontaneous magnetization of the Ising model},
  author={Baxter, Rodney J},
  journal={Journal of Statistical Physics},
  volume={145},
  number={3},
  pages={518--548},
  year={2011},
  publisher={Springer}
}

@article{wolff1989collective,
  title={Collective Monte Carlo updating for spin systems},
  author={Wolff, Ulli},
  journal={Physical Review Letters},
  volume={62},
  number={4},
  pages={361},
  year={1989},
  publisher={APS}
}

@article{metropolis1953equation,
  title={Equation of state calculations by fast computing machines},
  author={Metropolis, Nicholas and Rosenbluth, Arianna W and Rosenbluth, Marshall N and Teller, Augusta H and Teller, Edward},
  journal={The journal of chemical physics},
  volume={21},
  number={6},
  pages={1087--1092},
  year={1953},
  publisher={American Institute of Physics}
}

@article{yang1952spontaneous,
  title={The spontaneous magnetization of a two-dimensional Ising model},
  author={Yang, Chen Ning},
  journal={Physical Review},
  volume={85},
  number={5},
  pages={808},
  year={1952},
  publisher={APS}
}

@misc{github,
  author = {},
  title  = {},
  note   = {Simulation and image-processing code: \href{https://github.com/gabbiechang/knit-ising-model}{github.com/gabbiechang/knit-ising-model}}
}

@article{munden1959geometry,
  title={The geometry and dimensional properties of plain-knit fabrics},
  author={Munden, Dennis Lawrence},
  journal={Journal of the Textile Institute Transactions},
  volume={50},
  number={7},
  pages={T448--T471},
  year={1959},
  publisher={Taylor \& Francis}
}

@article{choi2003energy,
  title={An energy model of plain knitted fabric},
  author={Choi, Ka-Fai and Lo, Tien-Yu},
  journal={Textile Research Journal},
  volume={73},
  number={8},
  pages={739--748},
  year={2003},
  publisher={Sage Publications, Sage CA: Thousand Oaks, CA}
}

@misc{zenodo,
  author = {},
  title  = {},
  note   = {The swatches reported here and their high-resolution photographs are with archived onn the following Zenodo repository: \href{https://doi.org/10.5281/zenodo.23552182}{https://doi.org/10.5281/zenodo.23552182}.}
}

@article{cahn1958free,
    author = {Cahn, John W. and Hilliard, John E.},
    title = {Free Energy of a Nonuniform System. I. Interfacial Free Energy},
    journal = {The Journal of Chemical Physics},
    volume = {28},
    number = {2},
    pages = {258-267},
    year = {1958},
    month = {02},
    issn = {0021-9606},
    doi = {10.1063/1.1744102},
    url = {https://doi.org/10.1063/1.1744102}
}

@misc{matsumoto2026statistical,
  author = {Elisabetta A. Matsumoto},
  title  = {Statistical Knitting},
  note   = {$10^{\mathrm{th}}$ Edwards Symposyum, Cambridge, 26/10/2026.}
}

@article{hayama2025wave2weave,
  title={wave2weave: A Procedural Weave Data Generation},
  author={Hayama, Tatsuki and Uchibe, Kotaro},
  journal={Proceedings of the ACM on Computer Graphics and Interactive Techniques},
  volume={8},
  number={3},
  pages={1--10},
  year={2025},
  publisher={ACM, New York, NY}
}

@phdthesis{singal2024unraveling,
    author={Singal,Krishma},
    year={2024},
    school={Georgia Institute of Technology},
    title={Unraveling the Knot-So-Simple Behavior of Knitted Fabrics},
    pages={180},
    isbn={9798263396701},
}

\end{document}